# Venus as an Anchor Point for Planetary Habitability

Stephen R. Kane[1], Paul K. Byrne[2]

[1]Department of Earth and Planetary Sciences, University of California, Riverside, CA 92521, USA,

[2]Department of Earth, Environmental, and Planetary Sciences, Washington University in St. Louis, St. Louis, MO 63130, USA

**Abstract**

A major focus of the planetary science and astrobiology community is the understanding of planetary habitability, including the myriad factors that control the evolution and sustainability of temperate surface environments such as that of Earth. The few substantial terrestrial planetary atmospheres within the Solar System serve as a critical resource in studying these habitability factors, from which models can be constructed for application to extrasolar planets. The recent Astronomy and Astrophysics and Planetary Science and Astrobiology Decadal Surveys both emphasise the need for an improved understanding of planetary habitability as an essential goal within the context of astrobiology. The divergence in climate evolution of Venus and Earth provides a major, accessible basis for understanding how the habitability of large rocky worlds evolves with time and what conditions limit the boundaries of habitability. Here, we argue that Venus can be considered an "anchor point" for understanding planetary habitability within the context of terrestrial planet evolution. We discuss the major factors that have influenced the respective evolutionary pathways of Venus and Earth, how these factors might be weighted in their overall influence, and the measurements that will shed further light on their impacts of these worlds' histories. We further discuss the importance of Venus with respect to both of the recent decadal surveys, and how these community consensus reports can help shape the exploration of Venus in the coming decades.

# 1. Introduction

The topic of planetary habitability is one of the most active, exciting, yet challenging areas of active planetary and space science research. The importance of the topic largely relates to the astrobiological implications of understanding the sustainability of temperate surface conditions on planetary bodies but is also relevant to, for instance, forecasting Earth's future climate progression. In particular, the presence of surface liquid water throughout most of Earth's history is generally deemed as having played a crucial role in the origin and development of life on our planet, and may be a necessary—if on its own likely insufficient—requirement for similar biological processes elsewhere. Temperate surface conditions may have been present for a time on bodies before being truncated by planetary properties and/or environmental conditions, as in the case of Mars. Timescales that allow advanced biochemistry to emerge may require a relatively narrow range of planetary and stellar properties, and the primary objective of planetary habitability studies is thus to identify and quantify the boundaries of such parameters. Moreover, the predictive capabilities of a systems-based planetary science approach, accounting for interactions between interiors, atmospheres, and external factors, is crucial for inferring potentially habitable environments in the context of terrestrial exoplanet characterization.

The difficulty in assessing planetary habitability stems in large part from the limitations inherent to observing a single habitable planet at a particular epoch in its surface and atmospheric evolution. Although much may be inferred regarding Earth's past and future climate evolution, it remains challenging to study alternative pathways that could have rendered Earth's surface unable to harbour liquid water, or that might yet do so. The discovery of thousands of exoplanets, and the confirmation that terrestrial planets are among the most common types (Winn & Fabrycky 2015; Borucki 2016), provides a statistical framework for studying planetary properties and their evolution generally. Characterizing the surface environments of individual exoplanets remains enormously challenging and requires a model-based approach built on the foundation of Solar System observations and in-situ measurements (Horner et al. 2020; Kane et al. 2021; Kane 2022). Therefore, inferences of potentially habitable exoplanet environments are highly sensitive to our understanding of local planetary evolution pathways. Fortunately, the Solar System contains two large, rocky worlds where the present atmospheric and surface states differ substantially: Earth and Venus (Smrekar et al. 2018; Taylor et al. 2018; O'Rourke et al. 2023). The evolutionary pathway of Venus to its current runaway-greenhouse state is a matter of debate, having traditionally been attributed to its closer proximity to the Sun (Ingersoll 1969; Nakajima et al. 1992; Goldblatt & Watson 2012; Goldblatt et al. 2013). However, although Venus and Earth are of similar mass, size, and likely composition, there are numerous planetary-scale differences between them such as rotation rate, obliquity, and magnetic field. Thus, the precise contribution of insolation (incident solar radiation) flux to the overall evolution of Venus, including its atmosphere, surface, and interior, remains unclear. Venus is thus offers us a critical anchor point in the planetary habitability discourse, as its evolutionary story represents an alternate pathway from the Earth-based narrative —even though the origins of both worlds are, presumably, similar.

The recent National Academies' decadal surveys for astronomy and astrophysics ("*Pathways to Discovery in Astronomy and Astrophysics for the 2020s*", hereafter "*Astro2020*")[1] and for planetary science and astrobiology ("*Origins, Worlds, and Life: A Decadal Strategy for Planetary Science and Astrobiology 2023–2032*", hereafter "*OWL*")[2] both prioritise the understanding of habitable environments as a key research topic for the coming decades. The study of Venus, establishing its evolutionary pathway with respect to Earth, and recognising potential Venus surface environments inferred from exoplanet upper atmospheric spectra, will together form essential components of fulfilling this high cross-disciplinary science priority. Here, we discuss the study of

---

[1] https://www.nationalacademies.org/our-work/decadal-survey-on-astronomy-and-astrophysics-2020-astro2020
[2] https://www.nationalacademies.org/our-work/planetary-science-and-astrobiology-decadal-survey-2023-2032

Venus in the context of these decadal surveys' recommendations, and the required steps for realizing the astronomy and planetary science communities' goals using Venus as a planetary habitability anchor point.

**2. Recommendations of the Decadal Surveys**

The *Astro2020* Decadal Survey was publicly released on 4 November 2021, and identified three key priority areas, one of which was "Pathways to Habitable Worlds". Indeed, the words "habitable" and "habitability" feature 207 times in *Astro2020*. The proposed pathway to understanding planetary habitability mainly utilizes a strategy of exoplanet detection and characterization, including atmospheric measurements with current and planned facilities. However, the proposed strategy also emphasizes the critical need for basing the interpretation of exoplanet data on Solar System measurements and observations, thus requiring an important collaboration between the astronomy and planetary science communities. Venus is mentioned 15 times in *Astro2020*, mostly within the context of how the second planet represents an important, *local* example of terrestrial planetary evolution that can provide in-situ data and a framework for understanding the boundaries of terrestrial planetary habitability generally.

On 19 April 2022, the *OWL* Decadal Survey was published. *OWL* was structured into 12 overarching science questions, each a chapter, grouped under three themes: "Origins", "Worlds and Processes", and "Life and Habitability." These questions, in turn, were divided into sub-questions, each with their own corresponding "strategic research" (SR) areas and activities. Within *OWL*, the word "Venus" appears 261 times, and has particular prominence in numerous science questions spanning all three major themes in the report. Venus is included in 46 strategic research areas and activities in six priority science questions.

The role that Venus can play in our understanding of exoplanets and planetary habitability generally was given particular focus in *OWL*, such that the planet features in seven SRs in the twelfth, cross-cutting science chapter "Exoplanets." Strategic research activities in that chapter pertinent to Venus and exoplanet habitability include (but are not limited to): determining the properties of the atmospheres of terrestrial planets that would be observable on exoplanets to build a foundation for atmospheric characterization of analogue exoplanets (pg. 15-13); establishing where the inner edge of the Solar System's habitable zone is by studying the surface geomorphology and geochemistry of Venus to assess whether the planet ever possessed oceans (pg. 15-21); and studying methods to discriminate past and present false positive biosignatures on Solar System—such as abiotic $O_2$ on Venus—from true biosignatures we might detect on exoplanets (pg. 15-22).

The central role of Venus in understanding terrestrial planetary formation and evolution, planetary habitability, and establishing the likelihood of finding other Earth- or Venus-like exoplanets is emphasized time and again in both decadal surveys, demonstrating that Venus is recognised by the scientific community to be a singular destination at which these major planetary science problems can be tackled.

## 3. Fundamentals of Planetary Habitability

Planetary habitability is a foundational planetary and space science topic because identifying and understanding the factors that influence planetary surface conditions is essential for assessing whether planetary bodies might once have or do today host conditions amenable to life. However, defining precisely what planetary habitability means is a difficult and, at present, controversial topic. At its most basic level, one may consider habitability to mean "an environment capable of sustaining life". Though not inaccurate, this definition is sufficiently broad and vague so as to have limited practical application, particularly as definitions of life itself are not universally agreed upon (Cleland and Chyba 2002; Benner 2010). Even so, one commonly adopted method for producing a quantifiable planetary habitability framework is to assume a surface temperature range that allows liquid water to be present on an Earth-like world, which in turn defines the habitable zone (HZ) of a star (Kasting et al. 1993; Kopparapu et al. 2013, 2014; Kane et al. 2016; Hill et al. 2023). The advantage of this approach is that it is relatively agnostic to the development of life or its nature, whilst incorporating the dependence of Earth-based life on liquid water (Brack 1993). By this measure, investigations of planetary habitability can then focus on the conditions that allow surface liquid water to be sustained through geological time. Although intrinsically based on Earth life, the requirement for the presence of surface liquid water provides a necessary starting point from which to assess observational data.

In **Figure 1**, we show a graphical summary of the factors that can influence planetary habitability when seen through this framework. These properties are broadly divided into the influences of the star, the planetary system, and the intrinsic properties of the planet itself. Note that these factors vary enormously in the extent to which they influence a planet's energy budget, and the prevailing challenge for planetary scientists and astrobiologists is to robustly evaluate the relative contributions of each factor to planetary surface conditions. Yet a full consideration of all of these habitability factors is required for reliably modelling potentially habitable environments, including those present on exoplanets for which very limited data are available. Models of planetary habitability that can accurately match observations of Solar System worlds will be vital if we are to identify promising targets for follow-up observations with the finite resources available to study terrestrial exoplanetary atmospheres.

Given its protracted history of liquid surface water, our own planet has become the template for assessing potential habitability on planets around other stars. Yet observations of an exoplanet at a single epoch by definition cannot decode its full evolutionary history, and whether that world might once have had liquid water on its surface. Although Earth has largely maintained a habitable environment throughout its history, a critical lesson from our Solar System is that Earth-size planets can produce a substantial range of surface conditions. The inimicality of the Venusian surface compared with Earth may be the antithesis of habitability—but, since Venus illustrates the potential for non-habitability of Earth-size planets (and perhaps even a preview of the future of Earth itself), understanding the pathway to a Venus scenario is just as important as understanding the pathway to habitability that characterises Earth. **Table 1** provides a summary of some factors that govern planetary habitability, and the extent to which those factors are known for Venus and Earth.

*Astro2020* strongly recommended fully characterizing the divergence of the Venus–Earth habitability pathways, stating: "*Venus provided context for loss of habitability, with relevance for Venus-analogue extrasolar planets, and studies of stellar wind/planetary atmosphere interactions at Mars discovered and informed planetary atmospheric loss processes*" (pg. 288)

and

"*Combination of measurements and theory of the nature and processes that drove Earth's early habitability, and the loss of habitability on Venus and Mars can inform our understanding of exoplanet habitability*" (pg. 295).

The ability to study in-situ two Earth-size planets with dramatically different climate and habitability outcomes is an opportunity that exoplanets will never provide. Venus thus holds the key to understanding the conditions that enable long-term habitability in a manner that will allow reliable predictions of terrestrial planetary evolution generally.

**4. Divergence in Terrestrial Planetary Evolution: Earth and Venus**

In terms of bulk properties, Venus and Earth are remarkably similar: they are very close in size, and the mass of Venus is around 80% that of Earth (**Figure 2**). Yet there are major differences between the two planets, too. The insolation at Venus is almost twice that of Earth; its solid-body (retrograde) rotational period is 243 days; and the Venus atmosphere is almost entirely $CO_2$ (with a small amount of $N_2$ and trace abundances of other gasses such as $SO_2$, Ar, and water vapor). Moreover, the planet is cloaked in a global layer of $H_2SO_4$ clouds, which help give the planet a Bond albedo more than twice that of Earth. Together, the atmosphere's physical and chemical properties render the surface hotter than a self-cleaning oven with a pressure about that of 900 m underwater on Earth. And Venus seemingly lacks a contemporary intrinsic magnetic field—although whether remanent magnetism remains locked in the planet's surface rocks remains unknown (O'Rourke et al., 2019).

Measurements of the ratio of deuterium to hydrogen in the planet's atmosphere by the Pioneer Venus probe in the 1970s suggested that Venus had lost a considerable volume of water (Donahue et al., 1982). Reconciling that finding with Venus' contemporary surface conditions led scientists to form a view that the planet, forming with a steam-rich atmosphere above a magma ocean (e.g., Elkins-Tanton, 2008), would have been too close to the Sun to have been able to shed its heat to space other than by atmospheric escape of that water (e.g., Lebrun et al., 2013). If so, then the planet acquired its bulky atmosphere and hellish surface conditions early on and was never habitable. If, on the other hand, Venus were able to sufficiently cool for its atmospheric water to condense onto the surface (Hamano et al., 2013), then clouds could have helped keep surface conditions clement even under a steadily more luminous Sun (Way et al., 2016). Under that scenario, some other process—such as several major, contemporaneous volcanic eruptions akin to large igneous province-forming events on Earth (Way and Del Genio, 2020)— pushed Venus into its present runaway greenhouse state, possibly within the last billion years (Way and Del Genio, 2020; Krissansen-Totton et al., 2021).

The loss of oceans, or a steam-rich atmosphere, via a runaway greenhouse can lead to a temporary increase in the abundance of atmospheric $O_2$ (Luger and Barnes, 2015)—an increase that could be interpreted as a potential biosignature if detected in the atmosphere of an exoplanet. This process is contingent on several factors including planetary mass and stellar age and type but illustrates the point that, when seen from another planetary system at some time in the past, Venus could have been deemed habitable when it was anything but. At the least, then, Venus holds a cautionary tale for interpretations of apparently oxygen-rich atmospheres. Deciphering the evolutionary history of Venus is therefore key to bounding estimates of the number of Earth-like versus Venus-like exoplanets, especially those close to the outer perimeter of the "Venus Zone" (Kane et al., 2014) (**Figure 3**).

**5. Decadal Survey Priority Measurements**

Given these uncertainties regarding Venus, its evolution, and implications for general terrestrial planet evolution, we propose a two-pronged approach that engages both intrinsic Venus science and the statistics provided by the "exoVenus" (exoplanet analogs to Venus) population.

**5.1. Intrinsic Venus Properties**

Although we lack the ability with presently available data to establish which model for Venus' evolution is correct (Section 4), there *are* intrinsic properties of the planet we can measure that would enable us to definitively answer this question. For example, assaying the noble gas elemental and isotopic compositions of the Venus atmosphere would place vital constraints on models of the planet's initial volatile inventory and its history of water loss (Gillmann et al. 2009; Kane et al., 2019)—the very focus of a strategic research activity in Q.4 "Impacts and Dynamics" in *OWL* (pg. 6-22).

Similarly, understanding the history and rate of volcanic activity on Venus would enable us to place estimates on models of the interior and rates of degassing (Byrne and Krishnamoorthy, 2022). In addition to offering further insight into its atmospheric evolution, such an understanding would provide critical information regarding the planet's rates and mechanisms of heat loss, with implications for Venus' ability to generate and maintain a magnetic field and the attendant effects on its atmosphere (e.g., Driscoll, 2018). Investigations of volcanism and the history of magnetism on Venus are the foci of research activities described in Q.5 "Solid Body Interiors and Surfaces" in *OWL* (pg. 8-9).

A major means by which a planet is thought to regulate its climate is through the carbonate–silicate cycle, in which carbon is drawn out of the atmosphere and down into the interior (e.g.,

Walker et al., 1981). On Earth, this drawdown is principally accomplished with plate tectonics. Whether this process ever operated on Venus remains unknown, although at least spatially limited subduction has been documented on the planet (Davaille et al., 2017). Importantly, water plays a major role in the development and operation of plate tectonics (e.g., Tikoo and Elkins-Tanton, 2017). Further, the enigmatic tesserae on Venus—highly deformed rocks found on the planet's two major highlands as well as in smaller exposures elsewhere across the surface, and which are generally thought to be the oldest preserved material on Venus (e.g., Barsukov et al., 1986)—have been proposed to be equivalent to felsic, continental crust on Earth, on the basis of infrared emissivity, structural analysis, and collocation with areas of thickened crust (Hashimoto et al., 2008; Romeo and Turcotte, 2008). Given that large volumes of felsic crust are thought to require the presence of substantial amounts of surface or near-surface water and a mechanism for recycling water into the mantle (e.g., Campbell and Taylor, 1983), such as plate tectonics, the tesserae could reflect a period of sustained habitability in Venus' past (e.g., Hashimoto et al., 2008; Way et al., 2016). For this reason, establishing whether surface liquid water existed for a considerable amount of time on Venus is a strategic research activity in Q.6 "Solid Body Atmospheres, Exospheres, Magnetospheres, and Climate Evolution" in *OWL* (pg. 9-9).

**5.2. ExoVenus Candidates**

A parallel approach to studying the intrinsic properties of Venus is a statistical analysis of the vast (and still rapidly growing) inventory of terrestrial exoplanets (Ostberg et al. 2023). At present, exoplanetary radii are primarily measured with the transit method, a technique that has contributed to the bulk of exoplanet discoveries via missions such as Kepler (Borucki et al. 2010) and the Transiting Exoplanet Survey Satellite (TESS; Ricker et al. 2015). Many of these analyses concentrated on the fraction of stars that harbour a terrestrial planet within the HZ, referred to as "eta-Earth" (Dressing & Charbonneau 2013; Kopparapu 2013; Bryson et al. 2021). However, the transit method has a considerable bias towards the detection of short-period planets (Kane & von

Braun 2008), and thus is better suited to discovering planets that potentially have Venus-like atmospheric properties rather than that of Earth, and testing the boundaries of habitability.

Kane et al. (2014) defined the "Venus Zone" (VZ) as a target selection tool for identifying terrestrial planets where the atmosphere could potentially be pushed into a runaway greenhouse so as to lead to surface conditions similar to those of Venus. **Figure 3** shows the VZ (red) and HZ (blue) for stars of different temperatures. The outer boundary of the VZ is the "Runaway Greenhouse" line, which is calculated with climate models of Earth's atmosphere. The inner boundary (red dashed line) is estimated on the basis of where the radiation from the star would cause complete atmospheric erosion. The pictures of Venus shown in this region represent planet candidates detected by the Kepler mission. Kane et al. (2014) calculated an occurrence rate of VZ terrestrial planets of 32% for low-mass stars, and 45% for Sun-like stars.

However, the positions of the HZ and VZ boundaries are testable hypotheses only, since runaway greenhouse could occur beyond a calculated boundary (Foley 2015). Furthermore, as noted above, the insolation flux of a planet is one in a plethora of factors that influence planetary climate and has not (yet, at least) been clearly established as the *dominant* factor in determining whether a planet enters a runaway greenhouse state early in its evolution. Indeed, the exoplanet-focused pathway toward testing these boundaries is strongly advocated by *Astro2020*:

"*One important component of [detecting biosignatures] will be the study of exoplanets over a wide range of masses, ages, stellar insolations, and compositions; even uninhabitable planets can provide clues to how atmospheres are formed and evolve, as the atmosphere of Venus helped the understanding of the history of Earth*" (pg. 45)

and

*"Future space coronagraphic observations of young Venus or Mars analogs, for example, could help confirm whether those planets had more Earth-like atmospheres in the past"* (pg. 47).

Exoplanets orbiting bright stars will provide ideal opportunities for transmission spectroscopy follow-up observations with *JWST* and other facilities (Ostberg & Kane, 2019). Key spectroscopic features, most particularly $CO_2$ absorption at 2.7 and 4.3 μm, may be distinguished from those present in an Earth-similar atmosphere through the extension of wavelength coverage into the UV, where ozone absorption is prevalent (Ehrenreich et al. 2012; Barstow et al. 2016). The detection of prominent biosignature *combinations*, such as $H_2O$ and $CH_4$ together, may identify an atmosphere more likely to harbour temperate surface conditions (Schwieterman et al. 2018).

A primary recommendation of *Astro2020* is the continued development of direct-imaging capabilities, including acquisition of emission spectra across a broad range of wavelengths via disk-integrated reflectance spectra. These observations will be able to measure important atmospheric abundances and Rayleigh-scattering features for terrestrial planets (Cowan & Strait 2013; Madhusudhan 2019). Furthermore, NIR observations can reveal the temperature properties of terrestrial exoplanet cloud layers, and possibly even surface emissivity via infrared observing windows, as has been achieved for the nightside of Venus (Bezard et al. 1990; Arney et al. 2014). It is through such planetary spectroscopic analyses that the challenge of distinguishing between possible Venus and Earth-like surface conditions might best be overcome.

## 6. Conclusions

Fully understanding how a terrestrial planet becomes habitable and remains so is a fundamental challenge for the planetary science and astrobiology community, given the diversity and complexity of intrinsic and extrinsic processes that contribute to sustain habitable conditions over geological and biological time scales. In the face of this challenge, it is imperative that the full range of terrestrial planet atmospheric evolution data within the Solar System be exploited. Although Venus represents a clear end-member of planetary habitability, its contributions to understanding the prevalence of long-term temperate surface conditions on large rocky worlds have yet to be fully realised. Upcoming missions to Venus, including NASA's VERITAS (Cascioli et al. 2021) and DAVINCI (Garvin et al. 2022), and ESA's EnVision spacecraft (Widemann et al. 2023), will begin to flesh out this understanding. But the recent *Astro2020* and *OWL* decadal surveys present a united front by calling on the community to learn everything we can about Venus' evolutionary history. We should answer that call.

**Acknowledgements**

S.R.K. acknowledges support from NASA grant 80NSSC21K1797, funded through the NASA Habitable Worlds Program. The results reported herein benefited from collaborations and/or information exchange under NASA's Nexus for Exoplanet System Science (NexSS) research coordination network, which is sponsored by NASA's Science Mission Directorate. P.K.B. acknowledges support from Washington University in St. Louis. This research made use of NASA's Astrophysics Data System.

**Competing interests**

The authors declare no competing interests.

**Author Contributions**

Both S.R.K. and P.K.B. conceived of the idea of this Perspective. S.R.K. led the writing and the production of figures for the paper. P.K.B. contributed to the writing and the production of figures for the paper.

**Additional information**

Correspondence should be addressed to Stephen R. Kane.

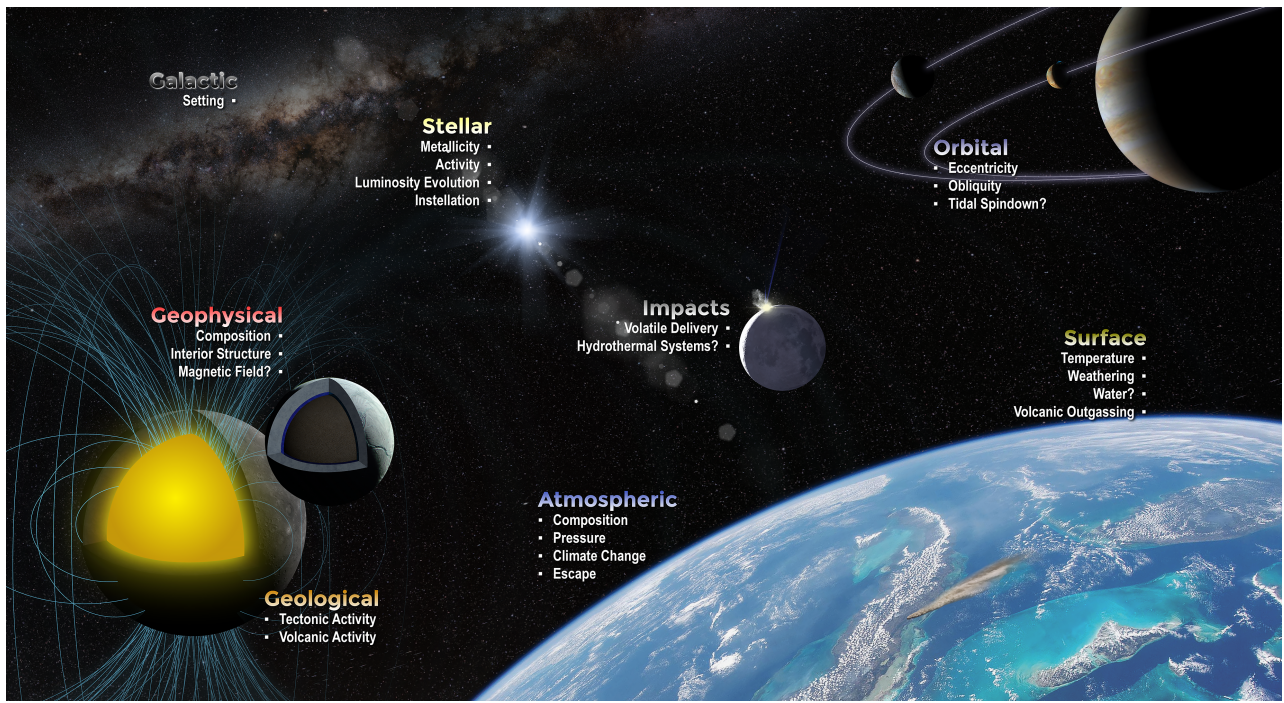

**Figure 1 | A graphical summary of the various factors that influence the surface conditions of a planet and their sustainability through time.** This illustration is pertinent to Venus, and includes the stellar (solar) radiation environment, the (Solar System's) make up and orbits of its major bodies, and intrinsic planetary properties such as geophysical, geological, atmospheric, and surface properties.

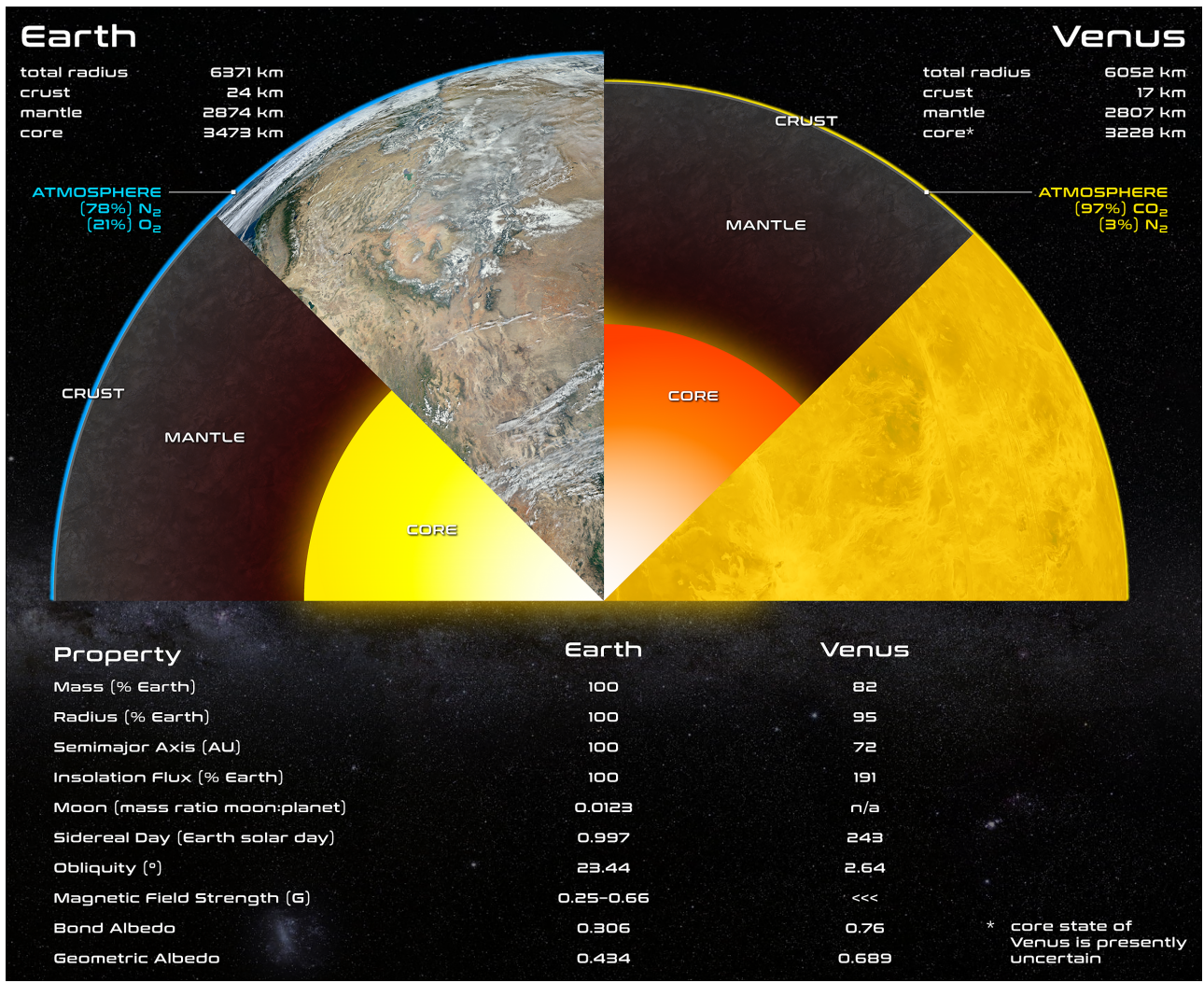

**Figure 2 | Schematic cross sections of Earth and Venus, showing the major internal components and atmospheric components, to scale.** For simplicity, oceanic and continental crust for Earth are not distinguished, nor is the interior structure of Earth's mantle shown. Note that there is considerable uncertainty regarding the state of the core of Venus, and so it is shown with orange fill instead of yellow. The interior structure for Earth is from Dziewonski and Anderson (1981). For Venus, the crust–mantle depth and mantle–core depth values are from James et al. (2013) and Aitta (2012), respectively. Under the cross sections, we show a table comparing key planetary properties of Earth and Venus (normalized to Earth values).

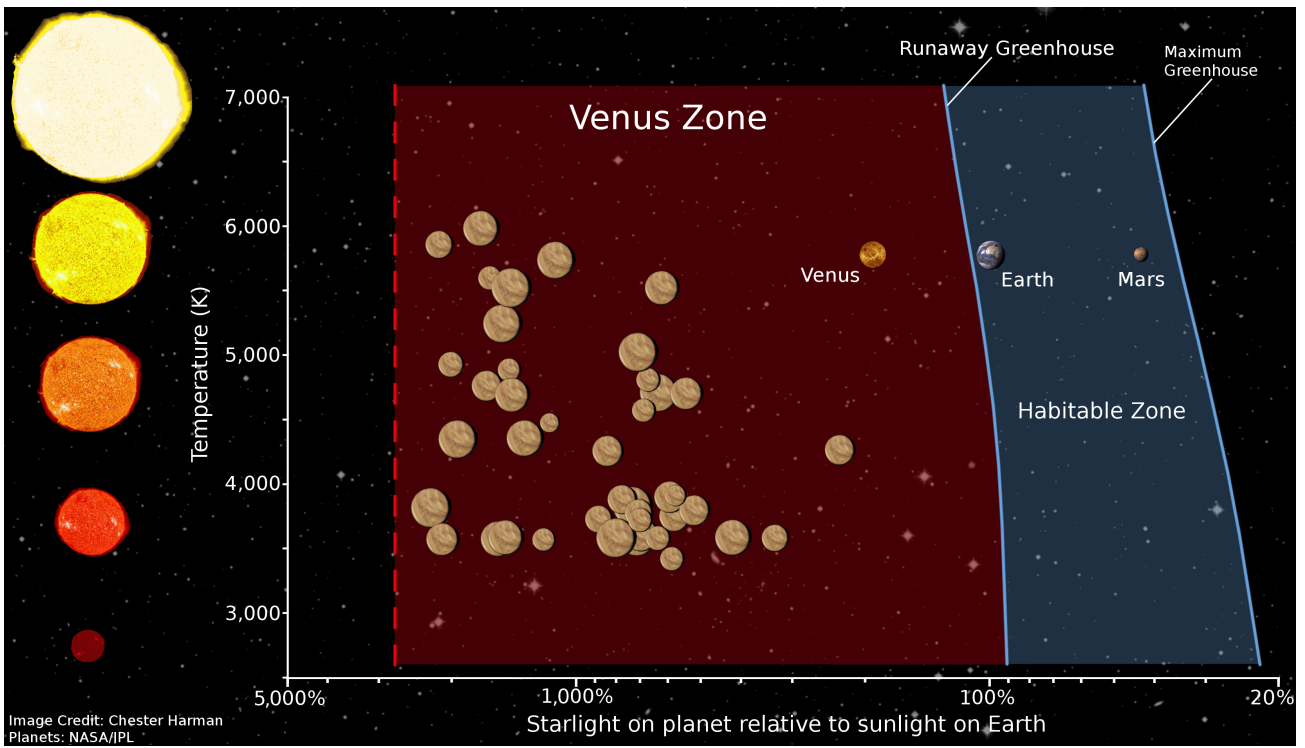

**Figure 3 | A representation of the Venus Zone and the Habitable Zone as a function of stellar effective temperature and insolation flux received by the planet.** The Venus Zone is shaded in red, with the Habitable Zone shown in blue. The images of Venus indicate the location of Kepler candidates that lie within the Venus Zone, scaled by the size of the planet. The Solar System planets of Venus, Earth, and Mars are also shown. Image credit: Habitable Zone Gallery/Chester Harman.

**Table 1 | The factors that govern planetary habitability, and the extent to which those factors are present for Earth and Venus.** The symbols have the following meanings. Y: Yes/Present; N: No/Absent; ?: Unknown/Uncertain; I: Insufficient Information. The superscript numbers are as follows. 1: the life-supporting elements carbon, hydrogen, nitrogen, oxygen, phosphorus, or sulphur (not all need be present); 2: interior heating is that energy derived from accretion, differentiation, radiogenic decay, and/or tidal dissipation; 3: the prospect for any element or molecule to be reduced or oxidized as a source of chemical energy for life; 4: substantial atmospheres only; exospheres (formed by, e.g., impact sputtering) are not included; and 5: intrinsically generated magnetic fields only.

|  |  | Earth | Venus |
|---|---|---|---|
| Water | Surface Liquid | Y | N |
|  | Subsurface Liquid | Y | N |
|  | Ground Ice | Y | N |
|  | Water Vapor | Y | Y |
| Chemistry | CHNOPS[1] | Y | I |
|  | Complex Organics | Y | I |
| Energy | Solar Heating | Y | Y |
|  | Interior Heating[2] | Y | Y |
|  | Redox[3] | Y | ? |
| Body | Atmosphere[4] | Y | Y |
|  | Magnetic Field[5] | Y | N |
|  | Present Habitability | Y | ? |
|  | Past Habitability | Y | ? |